# Cryptanalysis of "An Advanced Temporal Credential-Based Security Scheme with Mutual Authentication and Key Agreement for Wireless Sensor Networks"

Chandra Sekhar Vorugunti[1], Mrudula Sarvabhatla[2]

[1] Dhirubhai Ambani Institute of Information and Communication Technology,
Gandhi Nagar, Gujarat, 382007, India
vorugunti_chandra_sekhar@daiict.ac.in
[2] Sri Venkateswara University, Tirupathi, AP, 517502, India
mrudula.s911@gmail.com

**Abstract.** With the rapid advancement of wireless network technology, usage of WSN in real time applications like military, forest monitoring etc. found increasing. Generally WSN operate in an unattended environment and handles critical data. Authenticating the user trying to access the sensor memory is one of the critical requirements. Many researchers have proposed remote user authentication schemes focusing on various parameters. In 2013, Li et al. proposed a temporal-credential-based mutual authentication and key agreement scheme for WSNs. Li et al. claimed that their scheme is secure against all major cryptographic attacks and requires less computation cost due to usage of hash function instead encryption operations. Unfortunately, in this paper we will show that their scheme is vulnerable to offline password guessing attack, stolen smart card attack, leakage of password etc. and failure to provide data privacy.

**Keywords:** Wireless Sesnor Network, Cryptanalysis, key agreement; mutual authentication; temporal credential.

## 1  Introduction

Most of the real time applications in which WSN are deployed are data driven. The nodes need to sense and monitor various physical conditions like temperature, human presence etc. The data stored in these sensors are accessed through public channel Internet from remote areas. Hence authenticating the remote users trying to access the sensor data is more important. Many researchers have contributed into these literatures.

In 2013, Xue et al. proposed a temporal-credential-based mutual authentication and key agreement scheme which is secure and light weight. In 2013 Li et al. cryptanalyzed and shown that Xue at al. scheme is vulnerable to offline password guessing attack and proposed an improved version. Unfortunately, Li et al. scheme also suffers from offline and smart card lost attacks. In this paper we will show the drawbacks of Li et al. scheme.

## 2 Review of C.T Li et al. scheme

In this section, we examine the Li et al. [2] An Advanced Temporal Credential-Based Security Scheme with Mutual Authentication and Key Agreement for Wireless Sensor Networks. Li et al. Scheme is divided into the five phases of pre-registration, registration, login, mutual verification and session key agreement phases. The notations used in Li et al. scheme are listed below:

$U_i$: ith User.
$S_j$: jth Sensor node.
$ID_i$, $PW_i$: Identity and Password pair of user $U_i$
$SID_j$, $PW_j$: Pre-configured identity and password pair of the sensor node $S_j$
GWN: Gateway node
$K_{GWN\_U}/K_{GWN\_S}$ : Two private system parameters only know to GWN.
$TC_i$, $TC_j$:  A temporal credential issued by GWN to $U_i/S_j$
TS: The timestamp value.
$TE_i$ : The expiration time of $U_i$'s temporal credential.
SC: Smart card
$h(\cdot)$: A secure one-way hash function.
$A \oplus B$: Bit wise Exclusive OR (XOR) operation of A and B.
$A \| B$: Bit wise concatenation of A and B.

### 4.1. Pre-Registration Phase

Before registration GWN pre configures each user $U_i$ with a pair of identity $ID_i^{pre}$ and password $PW_i^{pre}$ . GWN stores $h(ID_i^{pre}\|PW_i^{pre})$, $ID_i^{pre}$ in its database. Similarly each sensor $S_j$ will be pre-configured with its identity $SID_j$ and a 160-bits random number $r_j$ and the value $h(SID_j\|r_j)$, Each $S_j$ has a pre-configured identity $SID_j$ and a 160-bits random number $r_j$ and the hash value of $S_j$'s pre-configured identity and random number $h(SID_j\|r_j)$ and $SID_j$ are stored on the GWN's side.

### 4.2. Registration Phase:

This phase has two parts for $U_i$ and $S_j$ and the details will be described as follows:
$U_i$ selects his/her own $ID_i$ and password $PW_i$. Then $U_i$ computes $VI_i = h(TS1\|H(ID_i^{pre}\|PW_i^{pre}))$, $CI_i=h(ID_i^{pre}\|PW_i^{pre})\oplus h(ID_i\|PW_i\|r_i)$, $DI_i = ID_i \oplus h(ID_i^{pre}\|PW_i^{pre})$ and sends $\{ID_i^{pre}, TS1, VI_i, CI_i, DI_i\}$ to GWN via an open and public channel, where TS1 is current timestamp value of $U_i$ and $r_i$ is a random number generated by $U_i$.
(U-2) After receiving the registration request from $U_i$, GWN checks if $|TS1-T^*GWN| < \Delta T$, where $T^*GWN$ is the current system timestamp of GWN and $\Delta T$ is the expected time interval for the transmission delay. If it does not hold, GWN sends REJ message back to $U_i$. Otherwise, GWN retrieves its own copy of $h(ID_i^{pre}\|PW_i^{pre})$ by using the parameter "$ID_i^{pre}$", computes $VI_i^* = h(TS1\| h(ID_i^{pre}\|PW_i^{pre}))$ and checks if $VI_i^* = VI_i$. If not, GWN terminates it; otherwise, GWN computes

$Q_i = CI_i \oplus h(ID_i^{pre} \| PW_i^{pre}) = h(ID_i \| PW_i \| r_i)$, $ID_i = DI_i \oplus H(ID_i^{pre} \| PW_i^{pre})$, $P_i = h(ID_i \| TE_i)$, $TC_i = h(K_{GWN\_U} \| P_i \| TE_i)$ and $PTC_i = TC_i \oplus Q_i$ and personalizes the smart card for $U_i$ with the parameters: $\{h(\bullet), h(Q_i), TE_i, PTC_i\}$. GWN maintains a write protected file, where the Status-bit indicates the status of the user, i.e., when $U_i$ is logged-in to GWN, the status-bit is set to one, otherwise it is set to zero. Finally, GWN sends $h(Q_i)$ and smart card to $U_i$ via an public and open environment.

(U-3) After receiving $h(Q_i)$ and smart card from GWN, $U_i$ checks whether the computed $h(h(ID_i \| PW_i \| r_i))$ is equal to $h(Q_i)$. If they are not equal, $U_i$ aborts this session and the smart card. Otherwise, GWN is authenticated by $U_i$. $U_i$ enters $r_i$ into his/her smart card and $U_i$'s smart card contains $\{h(\bullet), h(Q_i), TE_i, PTC_i, r_i\}$.

**Registration Phase of a Sensor**

(S-1) $S_j$ computes $VI_j = h(TS2 \| h(SID_j \| r_j))$ and sends $\{SID_j, TS2, VI_j\}$ to GWN via an open and public channel, where TS2 is current timestamp value of $S_j$.

(S-2) After receiving the message from Sj, GWN checks if $|TS2 - T^*GWN| < \Delta T$, where $T^*GWN$ is the current system timestamp of GWN and $\Delta T$ is the expected time interval for the transmission delay. If it does not hold, GWN sends REJ message back to $S_j$. Otherwise, GWN retrieves its own copy of $h(SID_j \| r_j)$ by using the key "$SID_j$", computes $VI_j^* = h(TS2 \| h(SID_j \| r_j))$ and checks if $VI_j^* = VI_j$. If not, GWN terminates it; otherwise, GWN computes $TC_j = h(K_{GWN\_S} \| SID_j)$, $Q_j = h(TS3 \| h(SID_j \| r_j))$ and $REG_j = h(h(SID_j \| r_j) \| TS3) \oplus TC_j$ and sends $\{TS3, Q_j, REG_j\}$ to $S_j$.

S-3) After receiving the message from GWN, $S_j$ checks if $|TS3 - Tj^*| < \Delta T$, where $Tj^*$ is the current timestamp value of $S_j$. If not, $S_j$ terminates it. Otherwise, $S_j$ checks whether the computed $h(TS3 \| h(SID_j \| r_j))$ is equal to $Q_j$. If they are equal, $S_j$ computes its temporal credential $TC_j = REG_j \oplus h(h(SID_j) \| r_j \| TS_3)$ and stores it. Note that $S_j$ does not need to store $r_j$ after finishing the phase.

**Login and Authentication Phase**

(A-1) $U_i$ computes $DID_i = ID_i \oplus h(TC_i \| TS4)$, $C_i = h(h(ID_i \| PW_i \| r_i) \| TS4) \oplus TC_i)$ and $PKS_i = K_i \oplus h(TC_i \| TS4 \| "000")$ and $h(TC_i \| TS4)$.

(A-2) After receiving the message from $U_i$, GWN checks the validity of TS4. If TS4 is valid for the transmission delay, GWN computes $TC_i^* = h(K_{GWN\_U} \| P_i \| TE_i)$ and $ID_i = DID_i \oplus h(TC_i^* \| TS4)$ and retrieves $U_i$'s password-verifier of $Q_i =$

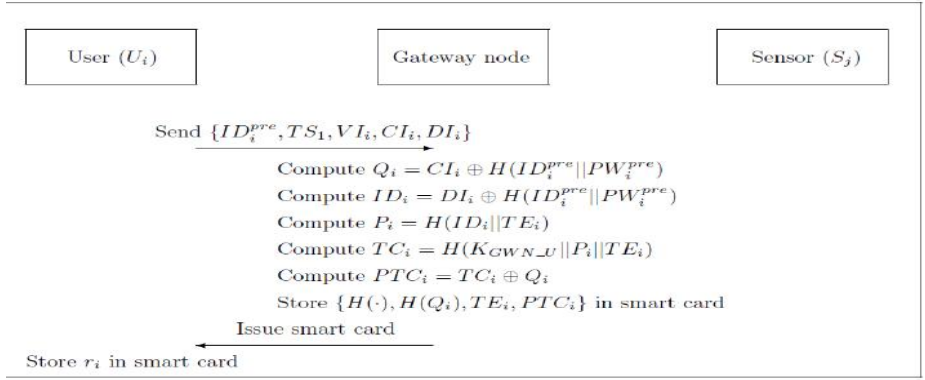

The above table taken from Li et al scheme [2].

$h(ID_i\|PW_i\|r_i)$ by using the parameter "$ID_i$". Then, GWN further computes $C_i^* = h(h(Q_i\|TS4)\oplus TC_i)$ and verifies whether $C_i^* = C^i$. If it does not hold, GWN rejects $U_i$'s login request; otherwise, the status-bit is set to one and TS4 is recorded in the 4th field of the identity table to demonstrate $U_i$'s last login. GWN computes $K_i = PKS_i \oplus h(TC_i\|TS4\|"000")$ and chooses a nearby suitable sensor node $S_j$ as the accessed sensor node. GWN further computes $S_j$'s temporal credential $TC_j = h(K_{GWN\_s}\|SID_j)$, $DID_{GWN} = ID_i \oplus h(DID_i\|TC_j\|TS5)$, $C_{GWN} = h(ID_i\|TC_j\|TS5)$ and $PKS_{GWN} = K_i \oplus h(TC_j\|TS5)$ and sends $\{TS5, DID_i, DID_{GWN}, C_{GWN}, PKS_{GWN}\}$ to $S_j$, where TS5 is current timestamp value of GWN.

(A-3) After receiving the message from GWN, $S_j$ checks the validity of TS5. If TS5 is valid for the transmission delay, $S_j$ computes $ID_i = DID_{GWN} \oplus h(DID_i\|TC_j\|TS5)$ and $C^*_{GWN} = h(ID_i\|TC_j\|TS5)$ and check if $C^*_{GWN} = C_{GWN}$. If not, $S_j$ terminates this session. Else, $S_j$ convinces that the received message is from a legitimate GWN. Moreover, $S_j$ computes $K_i = PKS_{GWN} \oplus h(TC_j\|TS5)$, $C_j = h(K_j\|ID_i\|SID_j\|TS6)$ and $PKS_j = K_j \oplus h(K_i\|TS6)$ and sends $\{SID_j, TS6, C_j, PKS_j\}$ to $U_i$ and GWN.

(A-4) After receiving the message from $S_j$, $U_i$ and GWN separately computes $K_j = PKS_j \oplus h(K_i\|TS6)$ and $C_j^* = h(K_j\|ID_i\|SID_j\|TS6)$. For GWN, if $C_j^* = C_j$, $S_j$ is authenticated by GWN. For the user $U_i$, if $C_j^* = C_j$, $S_j$ and GWN are authenticated by $U_i$. Finally, $U_i$ and $S_j$ can separately compute a common session key $KEY_{ij} = h(K_i \oplus K_j)$ and $U_i$ and $S_j$ will use $KEY_{ij}$ for securing communications in future.

## Cryptanalysis and Security Pitfalls of the C.T Li et al scheme

C.T Li cryptanalyzed recently proposed Xue et al.'s temporal-credential based mutual authentication scheme and proposed an improved scheme and claimed that their scheme requires less computation cost due to usage of lightweight one-way hash function and resists various cryptographic attacks like stolen verifier attacks, insider attacks, lost smart card attack and many logged-in users attack, etc. Unfortunately, In this paper we will show that the Li et al. scheme fails to resist all the attacks they claimed that their scheme resists and also the legal adversary can find out the password of the legal user.

## Offline Password Guessing Attack Coupled with Lost Smart Card Problem

Researchers [3,4] have shown that using various techniques like power consumption (The amount of power consumed by a microprocessor varies for operations like AES, DES, where as it is less for hash etc. ) analysis, Electromagnetic radiation, Timing etc. Many researchers including Li et al. [2] have cryptanalyzed the authentication schemes based on the above assumption. (Assumption that a smart card even though it is a tamper resistant, the data stored in its memory can be read out.)

In step (U-1) of registration phase of Li et al.'s scheme, $U_i$ sends $\{ID_i^{pre}, TS1, VI_i, CI_i, DI_i\}$ to GWN via an open and public environment, where TS1 is current time stamp value of $U_i$ and $VI_i = h(TS1 \| H(ID_i^{pre} \| PW_i^{pre}))$, $CI_i = h(ID_i^{pre} \| PW_i^{pre}) \oplus h(ID_i \| PW_i \| r_i)$, $DI_i = ID_i \oplus h(ID_i^{pre} \| PW_i^{pre})$. If an adversary $U_A$ intercepts the $U_i$'s registration message $\{ID_i^{pre}, TS1, VI_i, CI_i, DI_i\}$, by eavesdropping into the communication channel, $U_A$ can launch the off-line password guessing attack as follows:

Step 1: $U_A$ guesses a password $PW_i^{pre*}$ and computes $VIi^* = h(TS1 \| h(ID_i^{pre} \| PW_i^{pre*}))$,

Step 2: $U_A$ compares the result of $VIi^*$ with eavesdropped $VIi$.

A match in Step 2 above indicates the $PW_i^{pre}$ assigned by GWN to $U_i$ is $PW_i^{pre*}$. On getting $PW_i^{pre*}$, The adversary knows $ID_i^{pre}$, $PW_i^{pre*}$.

Step 3) $U_i$ smart card contains the values $\{h(\cdot), h(Q_i), TE_i, PTC_i, r_i\}$. If $U_A$ stolen the smart card of $U_i$ for a while , $U_A$ can perform following operations.

Step4) : $DI_i \oplus h(ID_i^{pre} \| PW_i^{pre}) = ID_i$

Step 5): $CI_i \oplus h(ID_i^{pre} \| PW_i^{pre}) = h(ID_i \| PW_i \| r_i)$

Step 6): $U_A$ can guess a password $PWi^*$ and computes $h(ID_i \| PW_i^* \| r_i)$ and compares the result with $DI_i \oplus h(ID_i^{pre} \| PW_i^{pre})$. If both are equal then the $U_i$ correct password is $PW_i^*$. On getting $U_i$ $ID_i$ and $PW_i$ the adversary UA can perform all major cryptographic attacks and can able to frame the session key also.